\documentclass[reprint,
superscriptaddress,
amsmath,amssymb,aps,showkeys,showpacs,
twoside,final,secnumarabic,
nofootinbib]{revtex4-2}

\usepackage[paperwidth=205mm,paperheight=290mm,top=17mm,bottom=25mm,
inner=17mm,outer=17mm,
twoside]{geometry}

\usepackage{cmap} 
\usepackage[T1,T2A]{fontenc}
\usepackage[utf8]{inputenc}
\usepackage[russian,main=english]{babel}
\usepackage{color}
\usepackage{graphicx}
\usepackage{dcolumn}
\usepackage{bm} 
\usepackage[unicode=true,colorlinks=true,linkcolor=magenta, urlcolor=blue, citecolor = blue,breaklinks]{hyperref}
\usepackage{multirow}
\usepackage{url}
\usepackage{breakurl}
\DeclareGraphicsExtensions{.eps}

\newcount\issue
\newcount\Vol
\newcount\numb
\usepackage{fancyhdr} 
\def\Vol{\textbf{79}}
\def\numb{x}
\begin{document}

\title{Machine Learning in Fundamental Physics
 \\[20pt] Application of Kolmogorov-Arnold Networks in high energy physics} 

\def\addressa{Skobeltsyn Institute of Nuclear Physics, Lomonosov Moscow State University, Leninskie gory, GSP-1, Moscow 119991, Russian Federation}

\author{\firstname{E.E.}~\surname{Abasov}}
\email[E-mail: ]{emil@abasov.ru}
\affiliation{\addressa}
\author{\firstname{P.V.}~\surname{Volkov}}
\affiliation{\addressa}
\author{\firstname{G.A.}~\surname{Vorotnikov}}
\affiliation{\addressa}
\author{\firstname{L.V.}~\surname{Dudko}}
\affiliation{\addressa}
\author{\firstname{A.D.}~\surname{Zaborenko}}
\affiliation{\addressa}
\author{\firstname{E.S.}~\surname{Iudin}}
\affiliation{\addressa}
\author{\firstname{A.A.}~\surname{Markina}}
\affiliation{\addressa}
\author{\firstname{M.A.}~\surname{Perfilov}}
\affiliation{\addressa}

\received{xx.10.2024}
\revised{xx.xx.2024}
\accepted{xx.xx.2024}

\begin{abstract}
Kolmogorov-Arnold Networks represent a recent advancement in machine learning, with the potential to outperform traditional perceptron-based neural networks across various domains as well as provide more interpretability with the use of symbolic formulas and pruning. This study explores the application of KANs to specific tasks in high-energy physics. We evaluate the performance of KANs in distinguishing multijet processes in proton-proton collisions and in reconstructing missing transverse momentum in events involving dark matter.
\end{abstract}



\maketitle
\thispagestyle{fancy}


\section{Introduction}\label{sec:intro}
Neural networks have been playing a significant role in high-energy physics for the past decades. They help improve event reconstruction in detectors, process classification, particle reconstruction, and many other important tasks. Potential architectures range from simple multi-layered perceptrons to complex transformer-based models~\cite{Plehn:2022ftl}. All of these types of networks, however, are based on the same principle of perceptron~\cite{mcculloch43a, rosenblatt1957perceptron}, which is, in essence, a learnable linear function with some fixed nonlinear activation function. The possibility of their usage for function fitting is based on the Universal Approximation Theorem~\cite{HORNIK1989359}.

Recently, a new approach was developed~\cite{liu2024kankolmogorovarnoldnetworks} that uses the Kolmogorov-Arnold representation theorem~\cite{Arnold1, Arnold2, Kolmogorov1, Kolmogorov2} instead. Kolmogorov-Arnold networks utilize learnable splines instead of linear functions and, thus, promise to provide greater level of accuracy then perceptron-based models. In addition, KANs allow for the splines to be fixed automatically or manually to a variety of analytical functions, which could prove instrumental in a greater understanding of neural networks output.


This study explores two potential applications for this new type of networks in the field of top quark physics: the separation of multijet processes, for which a traditional MLP is typically used, described in Section~\ref{sec:QCD}, and the reconstruction of the missing 4-momentum of invisible particles in processes involving dark matter, outlined in Section~\ref{sec:DM}. In addition, a comparison of the original implementation of KAN, further denoted as pyKAN~\cite{pyKAN}, and an optimized but feature-bare Efficient KAN (eKAN)~\cite{eKAN} is provided. All neural networks are constructed and trained in the PyTorch framework~\cite{paszke2019pytorchimperativestylehighperformance}.

\section{Classification task: multijet processes separation}\label{sec:QCD}
The first task on which KANs were tested is the separation of the multijet QCD background in the framework of single top-quark analysis. QCD processes have a very high production rate and represent an important, but difficult-to-model background. Therefore, in order to reduce bias from the QCD normalization uncertainty, this process is usually suppressed. 

Both the QCD background and other processes are modelled on the parton level in the POWHEG-BOX~\cite{Oleari:2010nx}, CompHEP4.6~\cite{CompHEP:2004qpa, Pukhov:1999gg} and MadGraph5~\cite{Alwall:2014hca, Frederix:2018nkq} Monte Carlo generators. The hadronization of these processes is done in Pythia8~\cite{Bierlich:2022pfr}, detector response is simulated in DELPHES~\cite{deFavereau:2013fsa} using the CMS detector profile.

\begin{figure}[ht]
    \centering
    \includegraphics[width=\linewidth]{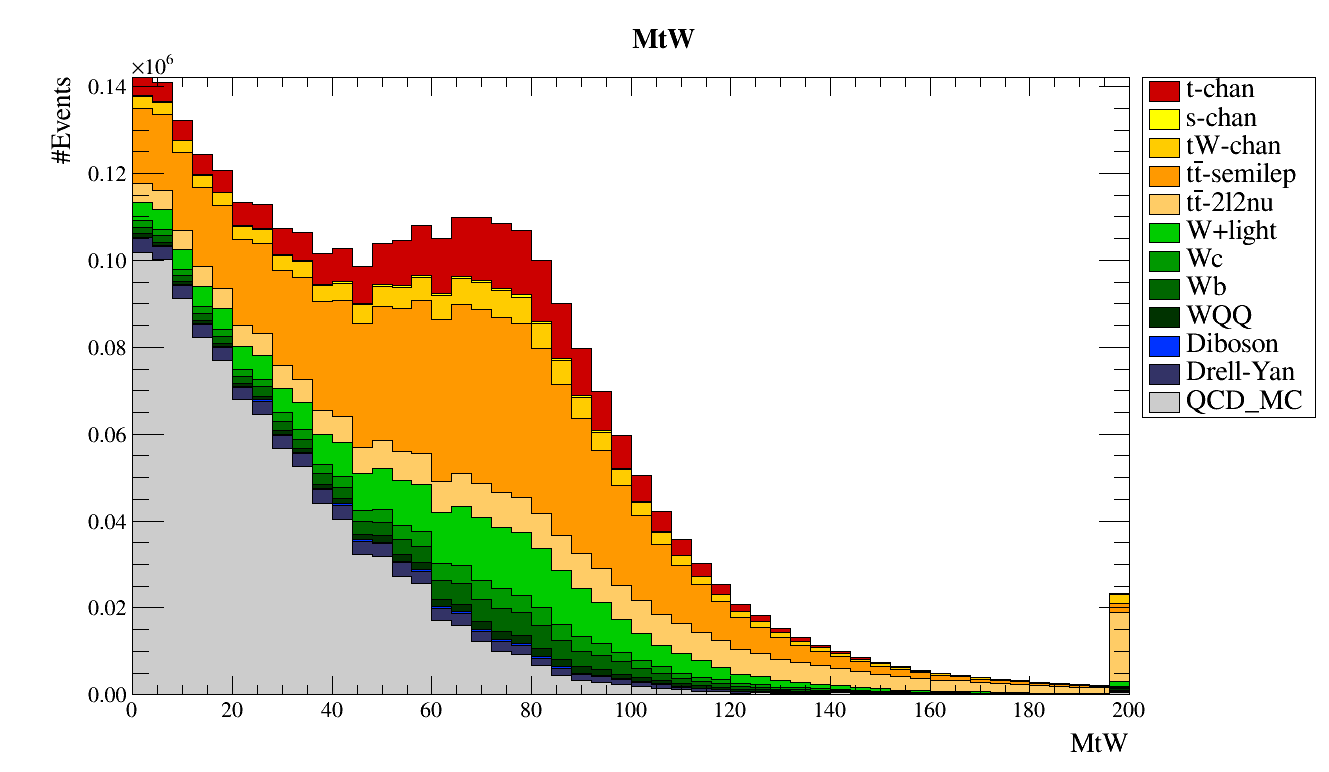}
    \caption{$M_{T}^W$ distribution for single-top production analysis. The QCD contribution is shown in grey.}
    \label{fig:MtW}
\end{figure}

The traditional approach of QCD suppression involves cuts on kinematic variables, sensitive to it, such as $M_{T}^W$, shown in Figure~\ref{fig:MtW}, a reconstructed transverse mass of the W boson from top-quark decay. However, this approach cuts out a substantial amount of other processes, too, and, as a consequence, reduces the statistics for the next steps of the analysis. 

\subsection{Baseline network}
\begin{figure}[ht]
    \centering
    \includegraphics[width=\linewidth]{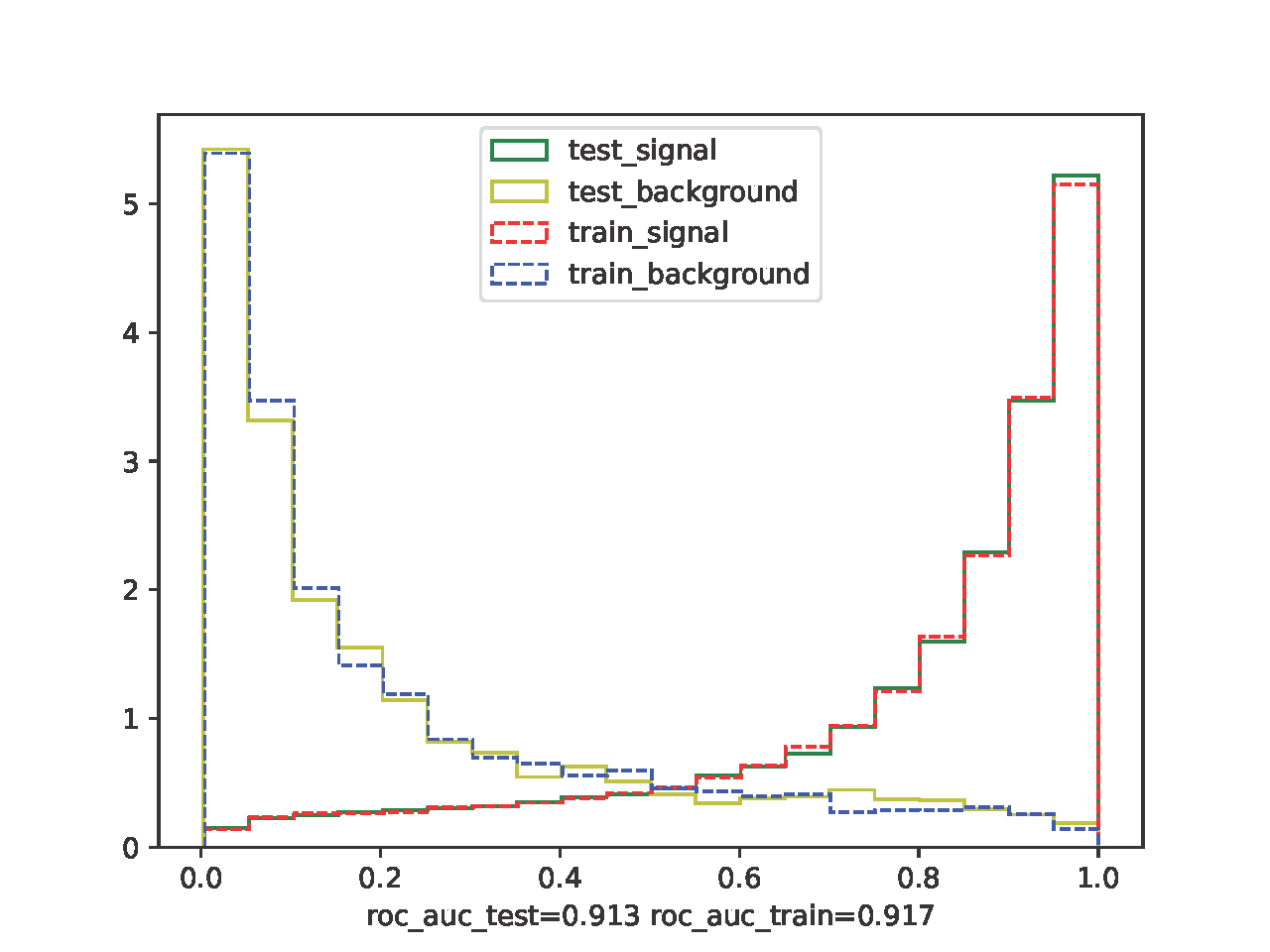}
    \includegraphics[width=\linewidth]{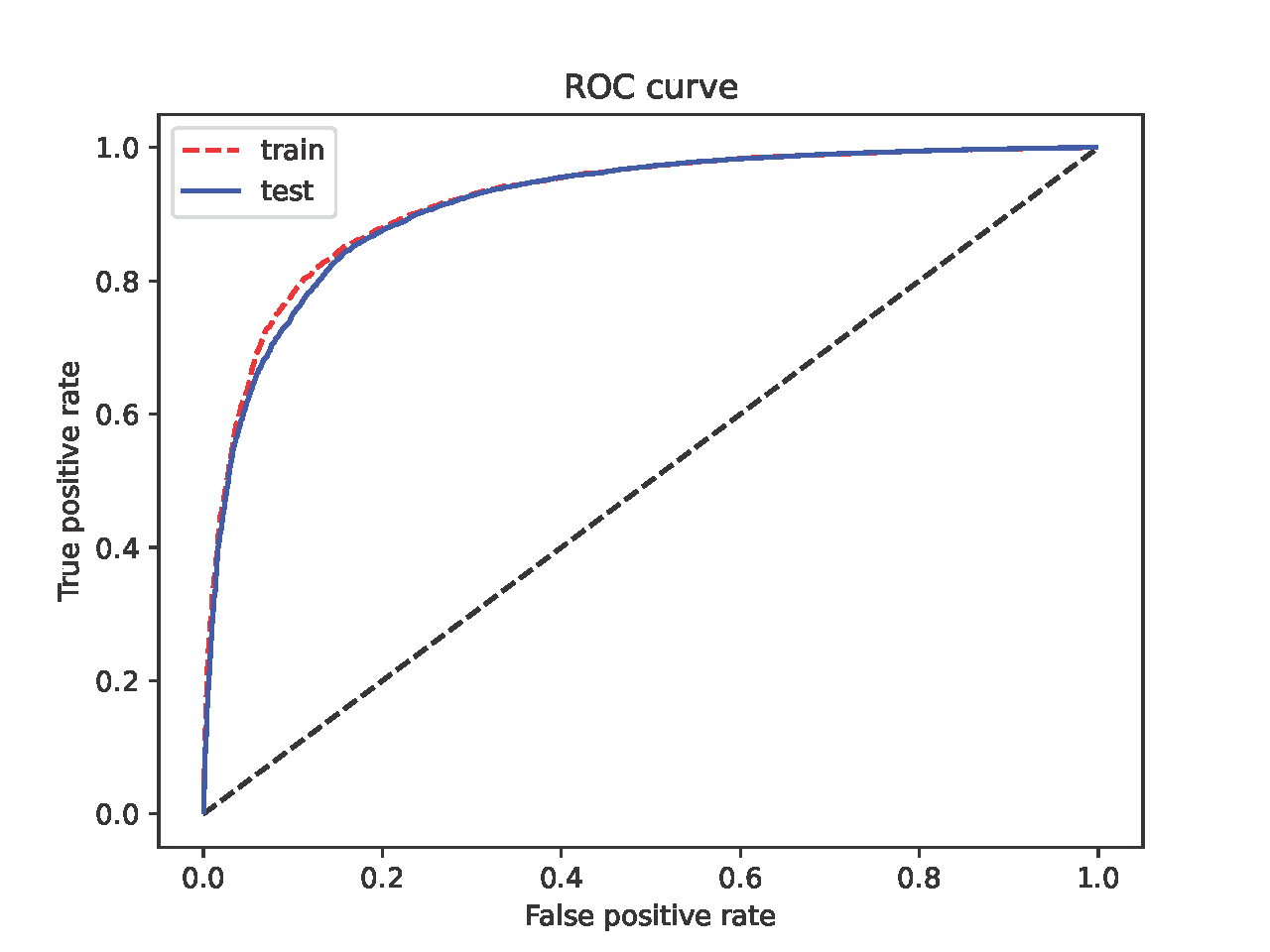}
    \caption{Output of the MLP for the multijet QCD separation (top) and the ROC curve (bottom).}
    \label{fig:QCD_MLP}
\end{figure}

Another approach to this problem, the use of neural networks, has proven to be more successful~\cite{abasov2023methodology638063678}. As an example of this method, a simple 2-layered perceptron with 100 neurons per layer is trained on the classification of QCD and other events, achieving the ROC AUC score of 0.91. LeakyRELU activation function is used, the network is trained using binary cross-entropy as the loss function. In order to prevent overfitting, dropout of 30\% is used in conjunction with L2 regularization. This MLP is taken as a baseline with which all KAN implementations are compared in this section, its performance is shown in Figure~\ref{fig:QCD_MLP}. 

\subsection{KAN implementation}
\begin{figure}[ht]
    \centering
    \includegraphics[width=.8\linewidth]{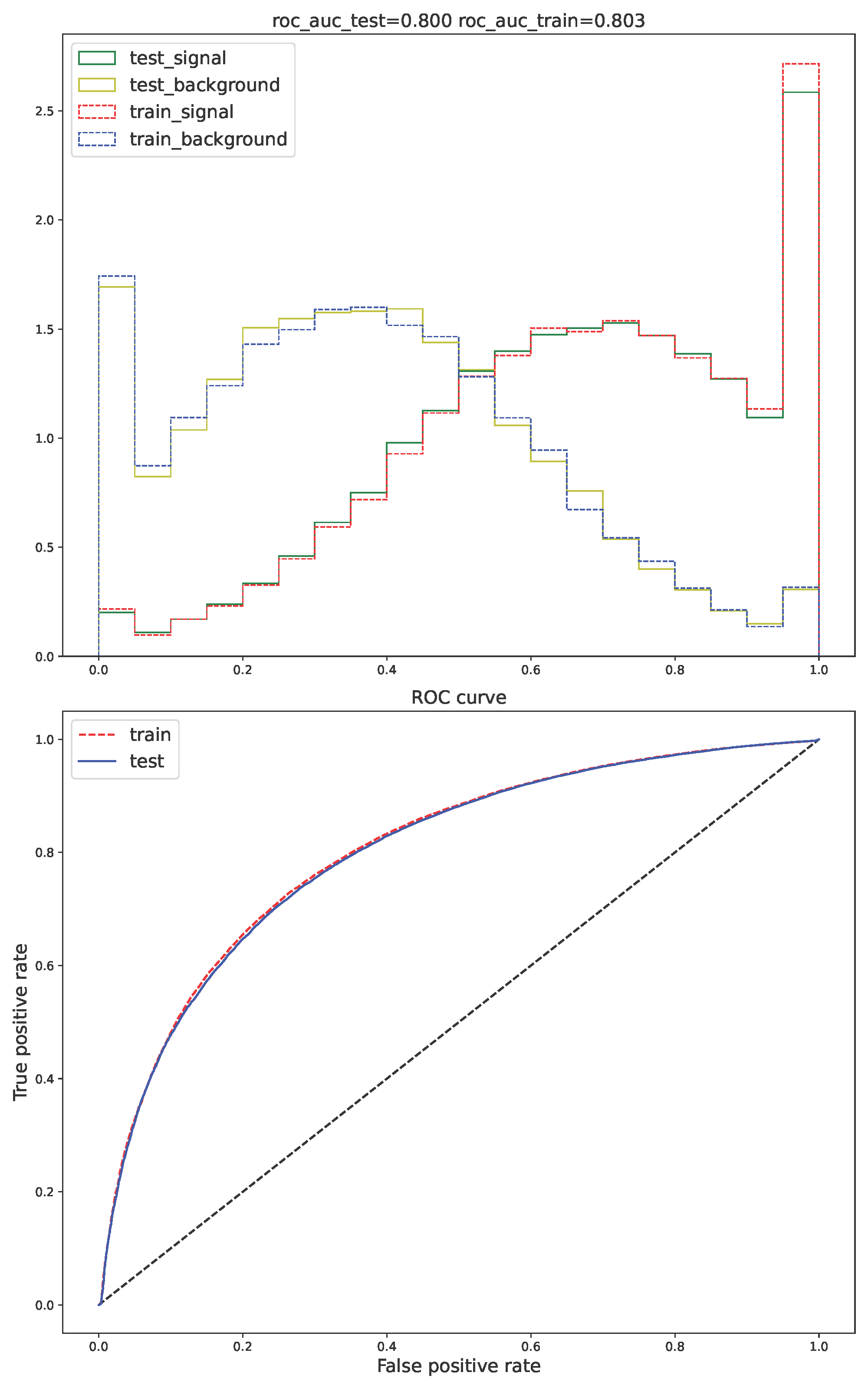}
    \caption{Output of the KAN for the multijet QCD separation (top) and the ROC curve (bottom). The network was trained on mean squared error as the loss function and clipped to the (0,1) range.}
    \label{fig:QCD_pyKAN}
\end{figure}

Firstly, the original pyKAN implementation~\cite{pyKAN} was tested. Despite the rich functionality, at the time of writing, it is plagued with various bugs, such as inability to train the network on GPUs, poor training with cross-entropy loss function, and general instability. Due to that fact, the only working solution for this task was to train the KAN using the mean squared error loss function instead of cross-entropy and then clip the output to the (0,1) range. Results were provided for two-layered KAN with 64 nodes in the middle layer. KANs utilize local B-spline basis functions, so the configuration of the grid, on which these functions are non-zero, is important. The amount of B-splines for each activation function is controlled by grid size. Cubic splines were used for the fitting, with a grid size of 5. It should be noted that the LBFGS optimizer was used for all KAN applications. As can be seen in Figure~\ref{fig:QCD_pyKAN}, the ROC AUC score, as well as the general shape of the distribution, is worse than the baseline MLP.

It should be noted that the pyKAN is being regularly updated with various bugfixes being implemented.

The ``Efficient KAN''~\cite{eKAN} is a pure bare-bones PyTorch implementation of KAN. The only functionality available in this library is the model initialization, without the ability to plot, prune, or use the symbolic regression. On the other hand, it does not exhibit the aberrant behaviour of the pyKAN.

\begin{figure}[ht]
    \centering
    \includegraphics[width=.8\linewidth]{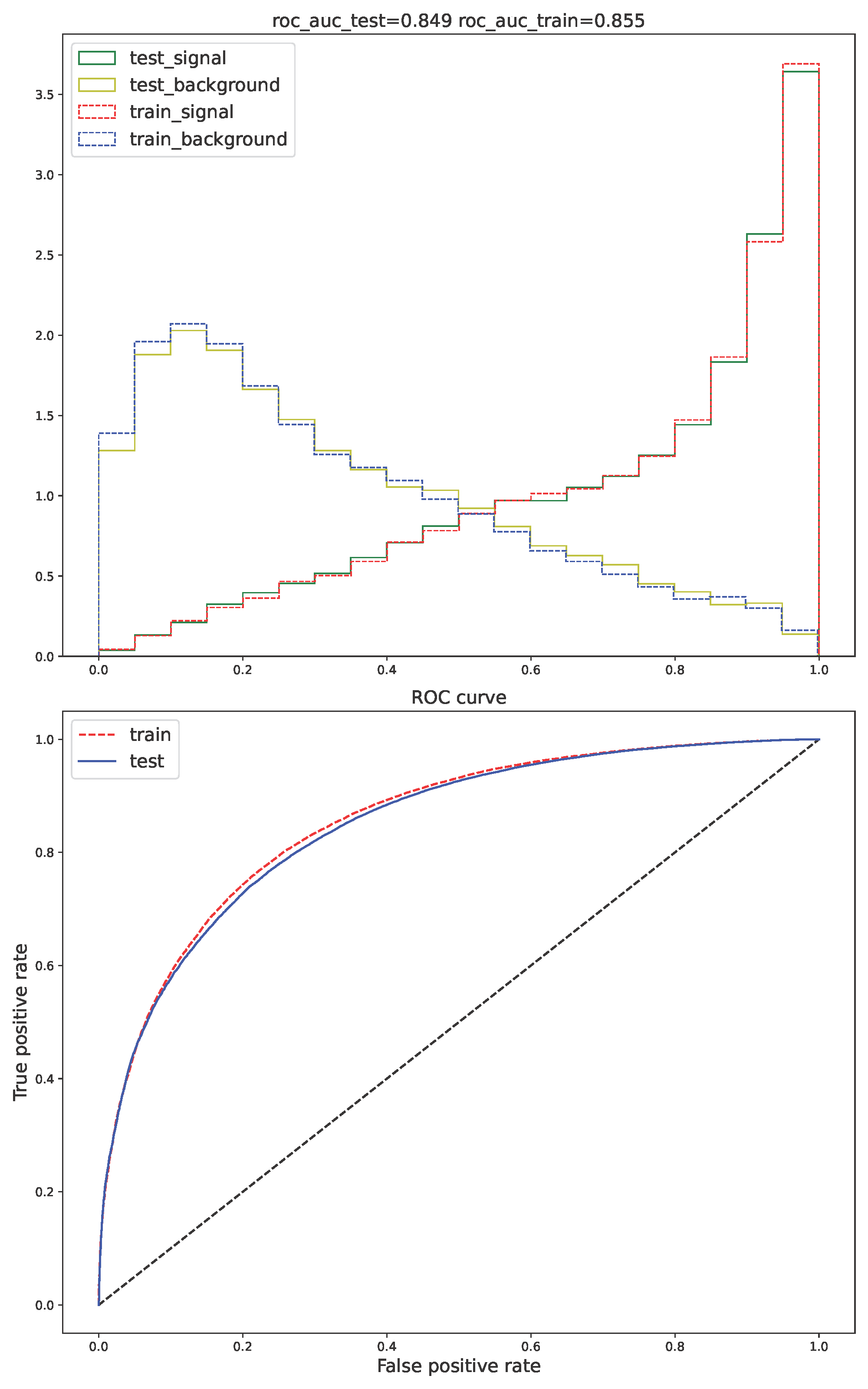}
    \caption{Output of the ``Efficient KAN'' for the multijet QCD separation (top) and the ROC curve (bottom). The network was trained with binary cross-entropy as the loss function.}
    \label{fig:QCD_eKAN}
\end{figure}

For the QCD classification, the shape of the network, as well as the spline order and grid are the same as in the pyKAN example. One noticeable difference in this case is the use of binary cross-entropy for the training procedure, which was not possible for pyKAN. The output of the network and its ROC curve are shown in Figure~\ref{fig:QCD_eKAN}. eKAN performs noticeably better than the pyKAN, mostly due to the availability of the cross-entropy as the loss function, while still not being able to beat the baseline.

In conclusion, in this task the baseline MLP proves to be superior to both KAN implementations, most likely owing to the simplicity of the task, which can be solved even with the simplest neural networks.

\section{Regression task: reconstruction of the missing momentum}\label{sec:DM}
Another possible application of Kolmogorov-Arnold networks arises in the study of top-quark production in association with dark matter particles. 

The possible interaction of dark matter particles with SM particles can be realized by means of an intermediate particle, the so-called mediator. For simplified DM models~\cite{Alwall:2008ag,Abdallah:2015ter,LHCNewPhysicsWorkingGroup:2011mji,Abercrombie:2015wmb} it is assumed that DM particles interact with SM particles by exchanging one or more particles, called ``mediators'', which have weak coupling with SM particles. This section focuses on single-top quark production with subsequent leptonic decay, which is well studied in the SM. One of the significant properties of such process in SM is correlation of the spin of top-quark and its decay products with the direction of the down-type quark. From that one can get the relation for the cosine of angle between down-type quark and the lepton in a top-quark decay. This relation changes in the presence of heavy scalar mediator, due to the change it makes in calculating the top-quark rest frame, in which the target variable is constructed~\cite{Abasov:2024nec}. Thus, in order to properly reconstruct this variable, it is required to separate contributions of two invisible particles: neutrino and dark matter mediator. 

\begin{figure}[ht]
    \centering
    \includegraphics[width=\linewidth]{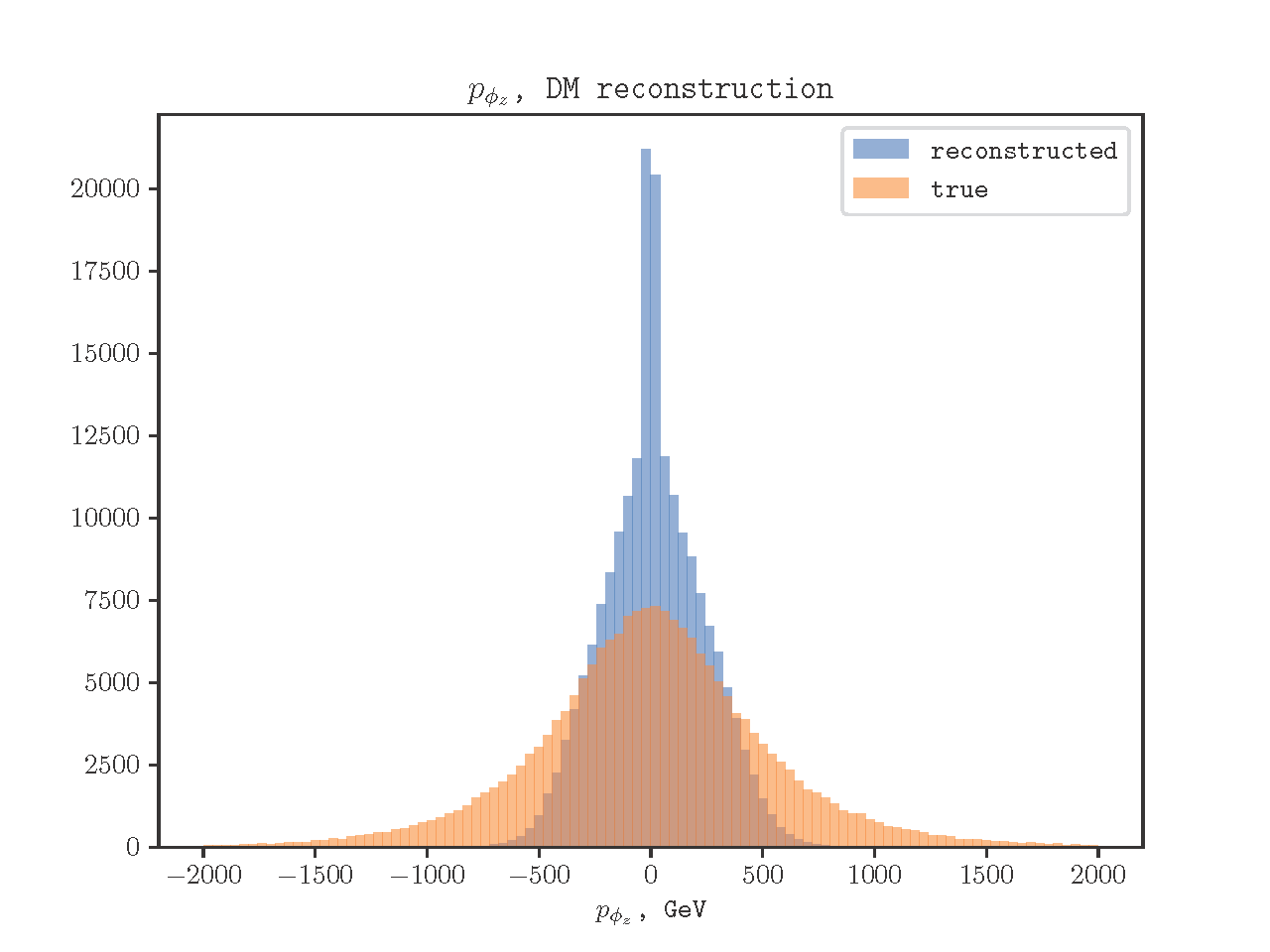}
    \includegraphics[width=\linewidth]{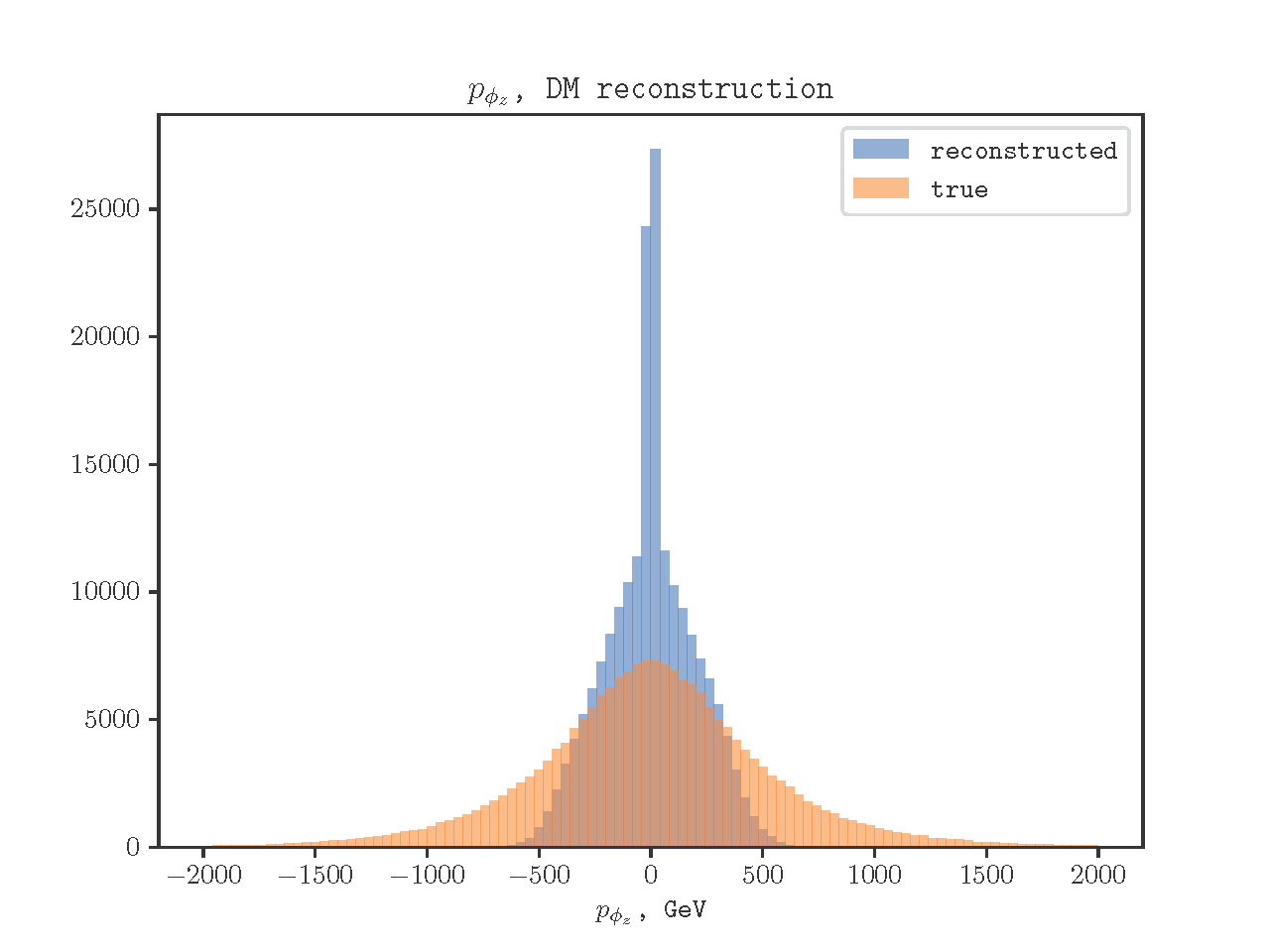}
    \caption{Reconstruction of the longitudinal component of the dark matter mediator momentum for the Baseline MLP from~\cite{Abasov:2024nec} (top), ``Efficient KAN'' (bottom).}
    \label{fig:DM_recon}
\end{figure}

\begin{figure}[ht]
    \centering
    \includegraphics[width=\linewidth]{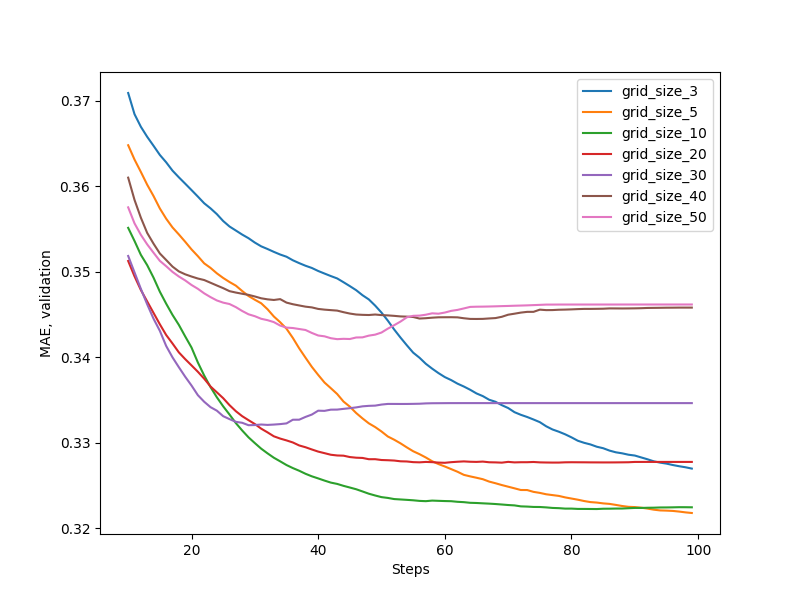}
    \includegraphics[width=\linewidth]{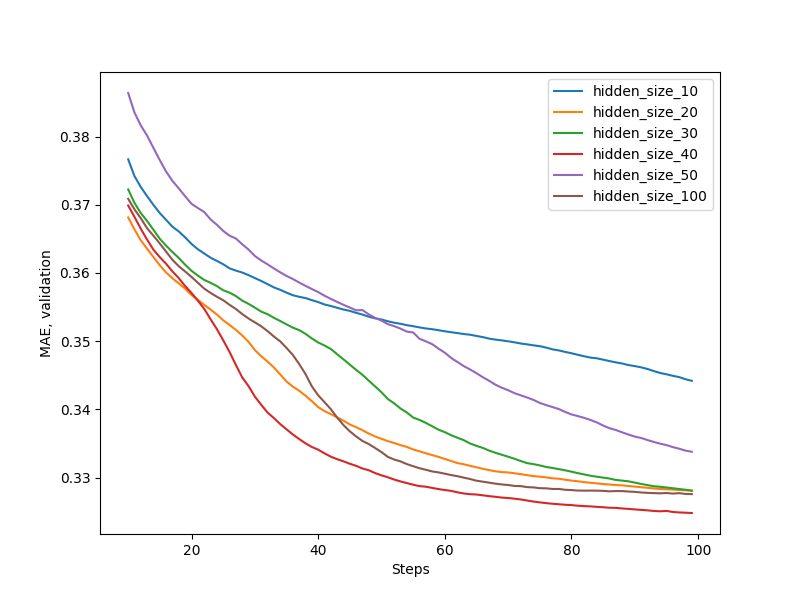}
    \caption{Hyperparameter optimization: grid size - top, hidden layer size - bottom. Training is shown from the 10th step.}
    \label{fig:DM_opt}
\end{figure}

As the baseline, the Multi-layered perceptron was used to reconstruct components of the 4-momentum of both particles, however, for the mediator this method did not yield desired results. Based on the same data, the Kolmogorov-Arnold network from the ``Efficient KAN'' framework is also trained. In order to compare the results obtained with the MLP baseline from~\cite{Abasov:2024nec}, the KAN is trained on the same objective function: Mean Absolute Error (MAE). The MAE score on the test dataset is 0.28 for the MLP and 0.30 for the KAN, which implies similar performance for these networks. This can be further reinforced by comparing the reconstructed momentum with its target value for both approaches (Figure~\ref{fig:DM_recon}). 
The KAN was further tuned with respect to the following hyperparameters: grid size, hidden size and depth of the network (Figure~\ref{fig:DM_opt}). The best performance was achieved with 2 hidden layers with 40 nodes per layer. With the performance of the neutrino and mediator momenta for KAN and MLP being almost identical, neither approach allows for adequate reconstruction of both particles.
\section{Results}\label{sec:Res}
Kolmogorov-Arnold networks provide a promising new alternative to traditional perceptron-based models. At the time of writing this article neither of available KAN implementations reaches full potential, with original pyKAN suffering from numerous bugs and the ``Efficient KAN'' and its derivatives lacking most of the original functionality, such as pruning and symbolic regression. For both of the problems discussed in this article, KANs perform either on par or slightly worse than traditional MLPs. With further improvements to the original pyKAN package, however, it is hoped that KANs could outperform traditional neural networks in these tasks and provide new level of interpretability.

During the publication process of the article, a new study on KANs in HEP for the task of $t\Bar{t}H / tH$ separation was published~\cite{Erdmann:2024unt}, which could provide further insight in this field.
\section*{FUNDING}
This work is conducted with financial support of grant RSCF 22-12-00152.

\clearpage


\end{document}